\documentclass[seceq,preprint]{ptptex}

\usepackage{graphicx}

\newcommand{\nn}{\nonumber\\}
\preprintnumber[3cm]{
RIKEN-TH-152\\ April, 2009}

\title{
Vacuum Structure Around Identity-Based Solutions
}

\author{
Isao \textsc{Kishimoto}$^1$ and Tomohiko \textsc{Takahashi}$^2$%
}

\inst{
$^1$Theoretical Physics Laboratory, RIKEN,
Wako 351-0198, Japan\\
\smallskip
$^2$Department of Physics, Nara Women's University,
Nara 630-8506, Japan
}

\recdate{April 21, 2009}

\abst{
We explore the vacuum structure in bosonic open string field theory
expanded around an identity-based solution parameterized by
$a\ (\geq -1/2)$. Analyzing the expanded theory using level
truncation 
approximation up to level 14, we find that the
theory has a stable vacuum solution for $a>-1/2$. The vacuum energy and
gauge invariant overlap numerically approach those of the
tachyon vacuum solution with increasing truncation level. 
We also find that, at $a=-1/2$, there exists an 
unstable vacuum solution in the expanded theory and it rapidly becomes
the trivial zero configuration just above $a=-1/2$. 
The numerical behavior of
the two gauge 
invariants suggests that the unstable solution corresponds to the
perturbative open string vacuum. These results reasonably support the
expectation that the identity-based solution is a trivial pure gauge
configuration for $a>-1/2$, but it can be regarded as the tachyon vacuum
solution at $a=-1/2$.
}

\begin{document}

\maketitle
\section{Introduction}

Bosonic open string field theory\cite{Witten:1985cc} (SFT) has classical
solutions describing a tachyon vacuum where D-branes with attached open
strings completely annihilate. A numerical tachyon vacuum solution was
first constructed using level truncation approximation in the Siegel
gauge.\cite{Kostelecky:1989nt,Sen:1999nx,Moeller:2000xv,Gaiotto:2002wy} 
Then, an analytic classical solution has been constructed
by Schnabl\cite{Schnabl:2005gv} and it has been found that the solution
possesses several properties of the tachyon vacuum. The vacuum energy of
the nontrivial analytic solution exactly cancels the D-brane
tension,\cite{Schnabl:2005gv,Okawa:2006vm,Fuchs:2006hw} and the
cohomology of the kinetic operator
around the vacuum is trivial.\cite{Ellwood:2006ba}
There are also several attempts to construct tachyon vacuum solutions
in superstring field theory as an extension of the bosonic
solution.\cite{Erler:2007xt,Aref'eva:2008ad,Fuchs:2008zx,Aref'eva:2009ac} 

In SFT, there is another type of analytic solution that was constructed
earlier on the basis of the identity string field instead of wedge
states used in Schnabl's
solution.\cite{Takahashi:2002ez,Takahashi:2001pp,Kishimoto:2005bs,
Kishimoto:2005wv,Kluson:2002ex,Kluson:2002gu,Kluson:2002te,
Kluson:2002av,Kluson:2003xu,Lechtenfeld:2002cu,Sakaguchi:2001kk}
An identity-based solution discussed in 
Refs.~\citen{Takahashi:2002ez,Kishimoto:2002xi,
Takahashi:2003ppa,Takahashi:2003xe} involves one parameter $a$ larger
than or equal to $-1/2$.
It is expected that the solution for $a>-1/2$ is a trivial pure gauge
solution because of the following facts:
\begin{enumerate}
 \item The solution can be expressed as a pure gauge form connecting
       to a trivial configuration\cite{Takahashi:2002ez}.
 \item The action obtained by expanding around the solution 
       can be transformed back to the action with the
       original BRST charge \cite{Takahashi:2002ez}.
 \item The new BRST charge gives rise to the cohomology,
       which has one-to-one correspondence to the cohomology of the
       original BRST charge \cite{Kishimoto:2002xi}.
 \item The expanded theory reproduces ordinary open string
       amplitudes\cite{Takahashi:2003xe}. 
\end{enumerate}
These are consistent with the expectation that the solution
corresponds to a trivial pure gauge.  On the other hand,
we find completely different properties in the
expanded theory around a solution with $a=-1/2$:
\begin{enumerate}
 \item[5.] The solution can be given as a type of singular gauge
	   transformation of the trivial
	   configuration\cite{Takahashi:2002ez}.
 \item[6.] The new BRST charge has vanishing cohomology in the
	   Hilbert space with the ghost number one\cite{Kishimoto:2002xi}.
 \item[7.] The open string scattering amplitudes vanish and the
	   result is consistent with the absence of open string
	   excitations\cite{Takahashi:2003xe}. 
\end{enumerate}
From these facts, it would be reasonable to expect that the
identity-based solution at $a=-1/2$ indeed corresponds to the tachyon
vacuum solution. 
Hence, we expect that the identity-based solution corresponds
to a trivial pure gauge form for almost all the parameter region
and it can be regarded as the tachyon vacuum solution
at $a=-1/2$.

The Schnabl solution can also be parameterized by $\lambda$.
The solution corresponds to a trivial pure gauge for
$-1\leq \lambda<1$.
However, at $\lambda=1$, 
it changes markedly to the tachyon vacuum solution.
Thus, the parameter $a$ in the identity-based solution has a property
similar to $\lambda$ in the Schnabl solution.
The similar dependence of these parameters suggests that the
identity-based solution may be gauge-equivalent to the Schnabl solution.

Unfortunately, we have not yet known how to
calculate the vacuum energy of the identity-based solution.
To calculate the vacuum energy, it is necessary to apply
a kind of regularization because a string field consists of an infinite
number of component fields. Indeed, the level can be regarded as
a regularization parameter for the numerical solution in the Siegel
gauge. Moreover, an analytic expression of the Schnabl solution includes
a parameter that regularizes the infinite sum of wedge-like states.
Hence, the difficulty of calculating the vacuum energy seems to
arise from the lack of such a regularization method for the
identity-based solution.  

However, we can provide indirect evidence that supports the possibility
of calculating the vacuum energy. The vacuum structure in the theory
expanded around the identity-based solution has been analyzed using
level truncation approximation and then we have found the following
results: 
\begin{enumerate}
 \item[8.] A numerical analysis shows that the nonperturbative vacuum
	   found for $a>-1/2$ disappears as $a$ approaches $-1/2$
	   \cite{Takahashi:2003ppa}.
 \item[9.] The energy of the nonperturbative vacuum for $a>-1/2$
          becomes closer to the value appropriate to cancel the D-brane
	   tension as the truncation level
       increases \cite{Takahashi:2003ppa}.
\end{enumerate}
These imply that the theory around the identity-based solution
for $a>-1/2$ has the tachyon vacuum, but the theory at $a=-1/2$ is
stable.
From consistency with the theory before expanding a string field,
it follows that the vacuum energy of the identity-based solution
itself is zero for $a>-1/2$ and it is equal to the
tachyon vacuum energy at $a=-1/2$.

The purpose of this paper is to perform additional numerical analysis of
vacuum structure in the theory expanded around the identity-based
solution and to provide further evidence for the expectation that the
identity-based solution can be regarded as the tachyon vacuum at
$a=-1/2$. The results of 8 and 9 are very
encouraging but they were based on a slightly lower level analysis.
First, we will raise the level from $(6,18)$ to $(14,42)$ searching the
tachyon vacuum in the expanded theory. As a result, we will numerically
confirm these results with higher precision.

As the parameter $a$ approaches $-1/2$, the tachyon vacuum solution
existing in the expanded theory for $a>-1/2$ annihilates into a trivial
zero configuration. From this fact, it is naturally expected that the
identity-based 
solution at $a=-1/2$ can be regarded as the tachyon vacuum solution.
If so, it is reasonable to expect that the expanded theory at
$a=-1/2$ has an unstable vacuum corresponding to the
perturbative open string vacuum in the original theory.
The unstable vacuum should emerge as the parameter $a$ approaches
$-1/2$. In this paper, we will find this emergence of the unstable vacuum
using level truncation approximation.

The paper is organized as follows.
In \S \ref{sec:annihilation}, after briefly reviewing the identity-based
solution, we will provide results for the annihilation of the stable
vacuum in the expanded theory. 
For each nontrivial solution, we will
calculate two gauge invariants: vacuum energy and gauge invariant
overlap. For $a>-1/2$, the former invariant should cancel the brane
tension and the latter should be nonzero, and both should
become trivially zero at $a=-1/2$.
We will see that this tendency becomes obvious as the truncation level
is increased. In \S \ref{sec:emergence}, we will consider the emergence
of an unstable vacuum. 
Indeed, using the level truncation up to $(16,48)$, 
we will find
that the unstable vacuum solution does exist in the expanded theory at
$a=-1/2$. Then, the vacuum energy for the solution approaches
the expected value with increasing level.
Moreover, we will show that the gauge invariant overlap is nearly
the expected value. In \S \ref{sec:summary}, we will give the summary and
discussion.

\section{Annihilation of tachyon vacuum
\label{sec:annihilation}}

The identity-based solution can be expressed
as\cite{Takahashi:2002ez,Kishimoto:2002xi,Takahashi:2003ppa,Takahashi:2003xe}
\begin{eqnarray}
\label{Eq:idsol}
 \Psi_0=Q_L(e^h-1)I-C_L((\partial h)^2e^h)I,
\end{eqnarray}
where $I$ is the identity string field associated with the star product,
and the half string operators $Q_L$ and $C_L$ are defined using the
BRST current $J_B(z)$ and ghost field $c(z)$ as follows,
\begin{eqnarray}
 Q_L(f)=\int_{C_{\rm left}} \frac{dz}{2\pi i} f(z)J_B(z),\ \ \ 
 C_L(g)=\int_{C_{\rm left}} \frac{dz}{2\pi i} g(z)c(z).
\end{eqnarray}
Here, $C_{\rm left}$ denotes the contour along a right semi-unit circle
from $-i$ to $i$, which conventionally corresponds to the left half
of strings. The function $h(z)$ is defined on the whole unit circle.
For the function $h(z)$ satisfying $h(-1/z)=h(z)$ and $h(\pm i)=0$,
the equations of motion, $Q_B\Psi_0+\Psi_0*\Psi_0=0$, hold
for the identity-based solution (\ref{Eq:idsol}).

Expanding the string field $\Psi$ around the solution $\Psi_0$ as
$\Psi=\Psi_0+\Phi$ and subtracting the vacuum energy 
at $\Psi_0$ from the original action,
 we can find the action for the fluctuation string
field $\Phi$:
\begin{eqnarray}
\label{eq:Q'action}
 S[\Phi]=-\frac{1}{g^2}\int \left(
\frac{1}{2}\Phi*Q'\Phi+\frac{1}{3}\Phi*\Phi*\Phi\right),
\end{eqnarray}
where the kinetic operator is given by
\begin{eqnarray}
 Q'=Q(e^h)-C((\partial h)^2 e^h).
\end{eqnarray}
The operators, $Q(f)$ and $C(g)$, are defined by
\begin{eqnarray}
  Q(f)=\oint \frac{dz}{2\pi i} f(z)J_B(z),\ \ \ 
  C(g)=\oint \frac{dz}{2\pi i} g(z)c(z),
\end{eqnarray}
where the integration contour is a unit circle.

Hereafter, we consider the expanded theory around the solution derived
from the function
\begin{eqnarray}
\label{Eq:hz}
 h(z)&=&\log\left(1+\frac{a}{2}\left(z+\frac{1}{z}\right)^2\right)\\
&=& -\log(1-Z(a))^2-\sum_{n=1}^\infty
\frac{(-1)^n}{n}Z(a)^n (z^{2n}+z^{-2n}),
\end{eqnarray}
where $Z(a)=(1+a-\sqrt{1+2a})/a$.
For the kinetic operator to be
well-defined, the parameter $a$ is larger than or equal to $-1/2$.
For this function, the kinetic operator can be
expanded as
\begin{eqnarray}
 Q'&=&(1+a)Q_B+\frac{a}{2}(Q_2+Q_{-2})+4aZ(a)c_0
-2aZ(a)^2(c_2+c_{-2})\nn
&&-2a(1-Z(a)^2)\sum_{n=2}^\infty(-1)^n
Z(a)^{n-1}(c_{2n}+c_{-2n}),
\end{eqnarray}
where we expand the BRST current and ghost field as
$J_B(z)=\sum_nQ_nz^{-n-1}$ and $c(z)=\sum_n c_n z^{-n+1}$,
respectively. 

The cohomology of this new BRST operator has been investigated in
Refs.~\citen{Takahashi:2002ez} and \citen{Kishimoto:2002xi}. 
For $a>-1/2$, the new BRST operator can be transformed to
the original BRST charge by a similarity transformation, and then the
cohomology has one-to-one correspondence to the original cohomology.
For $a=-1/2$, the new BRST operator has vanishing
cohomology in the Hilbert space with ghost number 1.
{}From these facts, it turns out that $a$ has 
properties similar to the parameter $\lambda$
 of the Schnabl solution.\footnote{The Schnabl
solution with a parameter $\lambda$ can be written as
\begin{eqnarray}
 \Psi=\frac{\lambda\partial_r}{\lambda\, e^{{\partial}_r}-1}\psi_r|_{r=0}.
\end{eqnarray}
}

We consider the tachyon vacuum in the expanded theory around
the identity-based solution. 
The expanded theory $S[\Phi]$ (\ref{eq:Q'action}) has a gauge symmetry
under
\begin{eqnarray}
\label{eq:gauge_tr}
 \delta\Phi=Q'\Lambda+\Phi*\Lambda-\Lambda*\Phi.
\end{eqnarray}
To find classical solutions in the theory,
we impose the Siegel gauge condition on the fluctuation string field;
$b_0\Phi=0$.
Under the Siegel
gauge condition, the potential can be expressed as
\begin{eqnarray}
\label{Eq:expaction}
 f_a(\Phi)=2\pi^2\left(
\frac{1}{2}\left<\Phi,c_0L(a)\Phi\right>
+\frac{1}{3}\left<\Phi,\Phi*\Phi\right>\right),
\end{eqnarray}
where it is normalized as $-1$ for the tachyon vacuum solution
at $a=0$.
Here, the operator $L(a)$ is given by
\begin{eqnarray}
\label{Eq:La}
 L(a)=(1+a)L_0+\frac{a}{2}(L_2+L_{-2})+a(q_2-q_{-2})
+4(1+a-\sqrt{1+2a}),
\end{eqnarray}
where $L_n$ is a total Virasoro generator and $q_n$ is a mode of the
ghost number current.\footnote{The expression is derived from
$L(a)=\{Q',\,b_0\}$. It can be rewritten only using ghost twisted
Virasoro operators as in Ref.~\citen{Takahashi:2003xe}.} 

\begin{figure}[t]
 \begin{center}
\includegraphics[width=12cm]{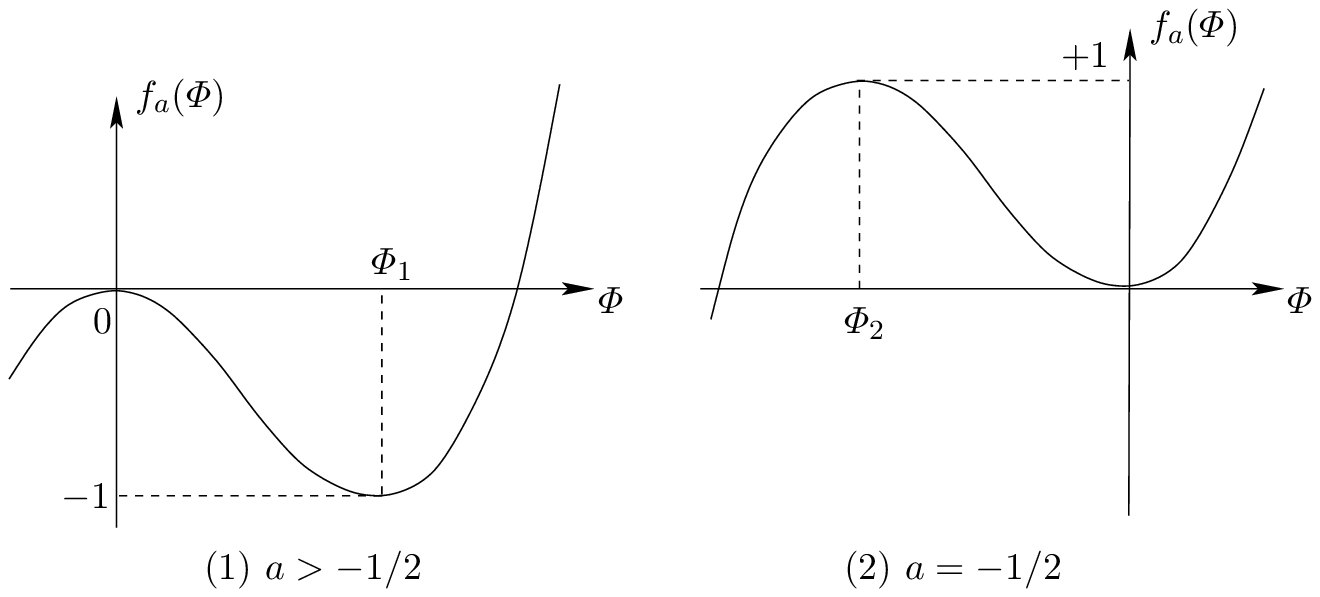}
\end{center}
\caption{Vacuum structure expected for the theory expanded around the
identity-based solution. (1) For $a>-1/2$, the theory should have a
 nontrivial vacuum solution, the vacuum energy of which cancels the
 D-brane tension. (2) At $a=-1/2$, the trivial configuration $\Phi=0$
 should be stable. 
An unstable solution is expected to exist and 
its vacuum energy should be equal to the D-brane tension.
}
\label{fig:vac}
\end{figure}

If the identity-based solution corresponds to a trivial pure gauge for
$a>-1/2$ and then to the tachyon vacuum for $a=-1/2$, the potential
(\ref{Eq:expaction}) should be illustrated as in Fig.~\ref{fig:vac}.
For $a>-1/2$, the tachyon vacuum configuration $\Phi_1$ should minimize
the potential, since the expanded theory for $\Phi$ is still
the theory on the perturbative open string vacuum. However, at $a=-1/2$, the
trivial 
configuration $\Phi=0$ should be stable since the expanded theory is
expected to be already on the tachyon vacuum. In addition, the theory at
$a=-1/2$ should have an unstable solution corresponding to the
perturbative open string vacuum of the unexpanded original theory. We
will discuss the unstable solution $\Phi_2$ in \S \ref{sec:emergence}.

\begin{figure}[t]
\begin{center}
\includegraphics[width=7cm]{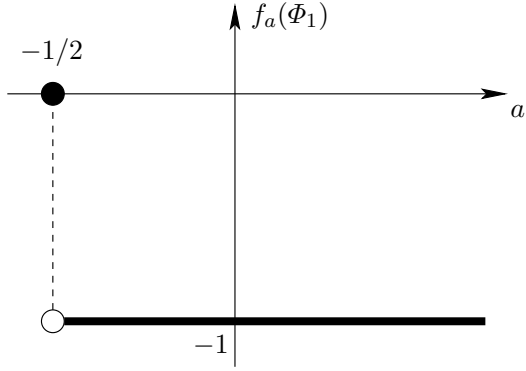}
\end{center}
\caption{Vacuum energy expected for the nontrivial stable solution
 $\Phi_1$.}
\label{fig:vacenergy1}
\end{figure}

Let us suppose that we obtain the tachyon vacuum solution $\Phi_1$ in
the expanded theory and we evaluate its vacuum energy.
Then, the vacuum energy $f_a(\Phi_1)$ is given as a function
of the parameter $a$. If the potential has such a structure as previously
expected, the vacuum energy is $-1$ for $a>-1/2$
but the solution $\Phi_1$ becomes trivial and therefore its vacuum energy
annihilates at $a=-1/2$ (see Fig.~\ref{fig:vacenergy1}).  

Now, let us consider the stable solution $\Phi_1$ by level
truncation calculation to confirm the above conjecture. Since the level
truncation is a good approximation, the vacuum energy for the
truncated solution is considered to approach the step function in
Fig.~\ref{fig:vacenergy1} as the truncation level is increased.
We apply an iterative approximation algorithm
as used in Refs.~\citen{Gaiotto:2002wy} and \citen{Kishimoto:2009cz} to
find the stable solution. 
Namely, with an appropriate initial configuration $\Phi^{(0)}$,
we solve
\begin{eqnarray}
\label{eq:iteration_La}
 c_0b_0(c_0L(a)\Phi^{(n+1)}+
\Phi^{(n)}*\Phi^{(n+1)}+\Phi^{(n+1)}*\Phi^{(n)}
-\Phi^{(n)}*\Phi^{(n)})=0,
\end{eqnarray}
$(n=0,1,2,\cdots)$ in the Siegel gauge.
If the iteration converges to a configuration $\Phi^{(\infty)}$,
it satisfies 
\begin{eqnarray}
c_0b_0(Q'\Phi^{(\infty)}+\Phi^{(\infty)}*\Phi^{(\infty)})=0,
\end{eqnarray}
which is a projected part of the equation of motion for $S[\Phi]$
(\ref{eq:Q'action}).
First, we construct a solution for $a=0$, namely, in the case of
$Q'=Q_B$, with the initial configuration:
\begin{eqnarray}
 \Phi^{(0)}=\frac{64}{81\sqrt{3}}\,c_1\left|0\right>,
\end{eqnarray}
which is the level zero solution.
In fact, the iteration using (\ref{eq:iteration_La}) converges
to the stable tachyon vacuum, which we denote as $\Phi_1|_{a=0}$.
Next, for $a=\epsilon$ ($0<|\epsilon|\ll 1$), 
we use the solution $\Phi_1|_{a=0}$ as the initial
configuration for the iteration  (\ref{eq:iteration_La}) 
 and obtain a solution  $\Phi_1|_{a=\epsilon}$.
Then, for $a=2\epsilon$, we use the solution $\Phi_1|_{a=\epsilon}$
 as the initial configuration and obtain a solution
 $\Phi_1|_{a=2\epsilon}$.
In this way, we can uniquely construct a solution
$\Phi_1|_{a=(n+1)\epsilon}$ from $\Phi_1|_{a=n\epsilon}$
$(n=0,1,2,\cdots)$.
In the case $\epsilon <0$, we find that
$\Phi_1|_{a=n_0\epsilon}$ reaches the trivial configuration,
namely, $\Phi_1|_{a=n_0\epsilon}=0$, for some $n_0$ such as
$n_0\epsilon > -1/2$.
Therefore, for $a\le n_0\epsilon$, 
the nontrivial stable solution $\Phi_1$ vanishes.

Actually, for each $a$, we continue the iterative
procedure until the relative error reaches $10^{-8}$
and we also check the BRST invariance\cite{Hata:2000bj} of the resulting
configuration,
$b_0c_0(Q'\Phi^{(\infty)}+\Phi^{(\infty)}*\Phi^{(\infty)})=0$,
 as in Ref.~\citen{Kishimoto:2009cz}.\footnote{Here, the BRST invariance
 that we checked is an invariance in the expanded theory. Namely, the BRST
transformation is given by $\delta_B\Phi=Q'\Phi+\Phi*\Phi$.}
For each calculation, we use the $(L,3L)$ truncation.
Namely, we truncate the string field 
to level $L\equiv L_0+1$ and interaction terms up to total level $3L$.

\begin{figure}[t]
 \begin{center}
\includegraphics[width=11.5cm]{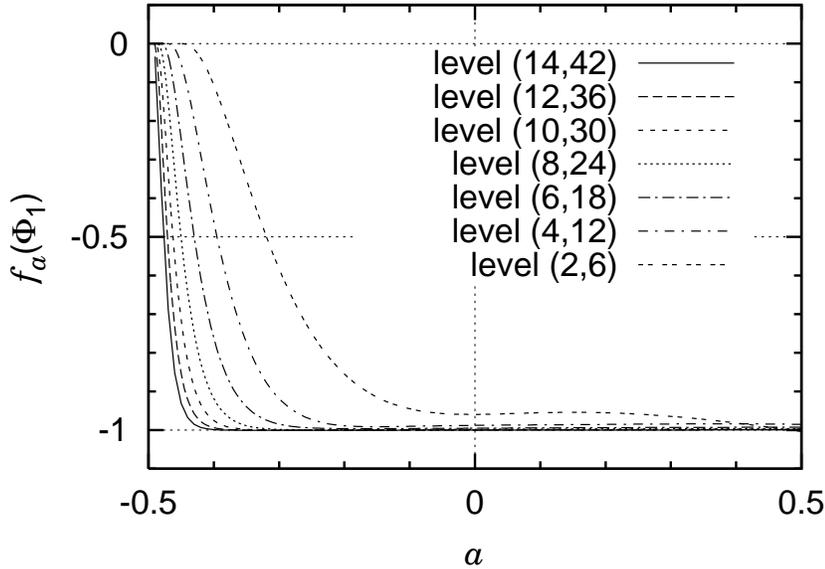}
\end{center}
\caption{Vacuum energy of the numerical stable solutions in the expanded
 theory around the identity-based solution. As the truncation level is
 increased, the resulting plots approach the step function expected
 as in Fig.~\ref{fig:vacenergy1}.} 
\label{fig:vacenergy2}
\end{figure}

First, we show plots of the vacuum energy for the resulting solution in
Fig.~\ref{fig:vacenergy2}. We can find that, for various $a$,
the resulting plots approach the tachyon vacuum energy $-1$
as the truncation level is increased. 
Particularly around $a=0$, the plots so rapidly approach $-1$ that they
are indistinguishable from
each other for higher levels than $(8,24)$. 
Then, for decreasing $a$ to $-1/2$, the vacuum energy increases
rapidly to zero from $-1$. As a whole, the plots become closer to the
step function 
as depicted in Fig.~\ref{fig:vacenergy1} as the truncation level
increases.
\begin{figure}[h]
 \begin{center}
\includegraphics[width=11.5cm]{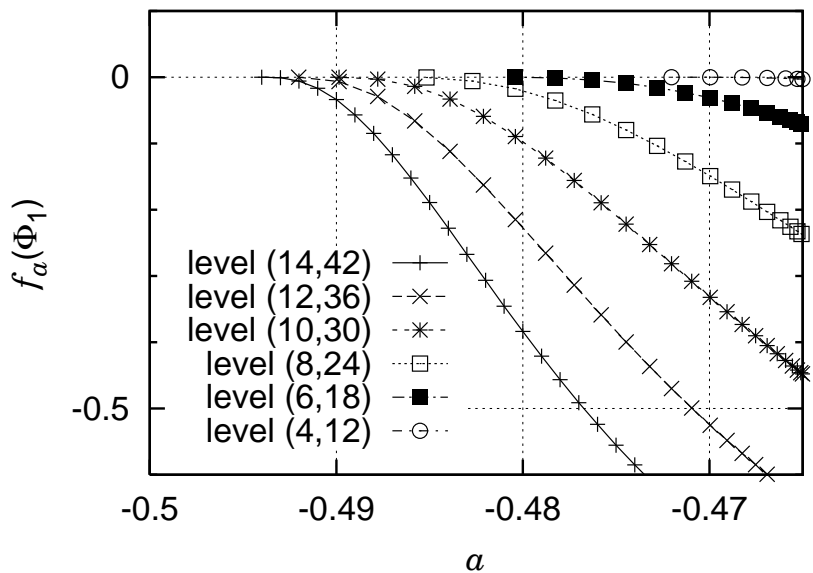}
\end{center}
\caption{Enlarged view of the vacuum energy for $\Phi_1$ around $a=-0.5$.}
\label{fig:vacenergy3}
\end{figure}
In Fig.~\ref{fig:vacenergy3},
we display an enlarged view of the vacuum energy around $a=-0.5$.
We find that at the point at which the vacuum energy
is zero, the numerical solution itself becomes trivially zero.
Namely, in the level truncated theory, the nontrivial stable solution
gradually annihilates as the parameter $a$ approaches a critical
value. Furthermore, this annihilation point becomes
closer to $a=-0.5$
as the level becomes larger. All these results support our expectation
for the vacuum structure associated with the stable solution $\Phi_1$.

Next, we consider the gauge invariant overlap for the numerical stable
solution. The gauge invariant overlap in the original theory is given as
\begin{eqnarray}
\label{eq:O_Psi}
 O(\Psi)&=&N\left<I\right|V\left(\pi/2\right)\left|\Psi\right>,
\end{eqnarray}
where $I$ denotes the identity string field, $V(\pi/2)$ is an onshell
closed string vertex operator inserted at a string midpoint, and $N$ is
a normalization
constant.\cite{Zwiebach:1992bw,Hashimoto:2001sm,Kawano:2008ry,Ellwood:2008jh} 
(Hereafter, we take the same normalization as that in
Ref.~\citen{Kishimoto:2009cz}.)
In the theory with the ordinary BRST
charge, the gauge invariant overlap is to be zero for a trivial pure
gauge configuration, but it becomes a nonzero value for the tachyon
vacuum solution. These facts have been confirmed by both analytical
and numerical calculations.\cite{Kawano:2008ry,Ellwood:2008jh}

The gauge invariant overlap $O(\Psi)$ (\ref{eq:O_Psi}) is also
left invariant under the gauge transformation (\ref{eq:gauge_tr})
in the expanded theory.\footnote{
Note that $\langle I|V(\pi/2)(c_n+(-1)^nc_{-n})=0$ and 
$\langle I|V(\pi/2)(Q_n+(-1)^nQ_{-n})=0$ hold.
}
Since the level truncation is applicable to the
evaluation of the invariant overlap in the original
theory,\cite{Kawano:2008ry} it is naturally 
expected that we 
can also calculate the overlap for the numerical stable solution in the
expanded theory. Contrasting to the vacuum energy, the invariant
overlap does not include explicitly the parameter $a$. However, the overlap
is given as the function of $a$ since the stable solution itself depends on
the parameter. If the expanded theory has the vacuum structure
expected above, the overlap should be nonzero for almost all $a$, but
it is to be zero at $a=-1/2$. This behavior is considered to be
approximately realized for the numerical stable solution. 
\begin{figure}[t]
 \begin{center}
\includegraphics[width=11.5cm]{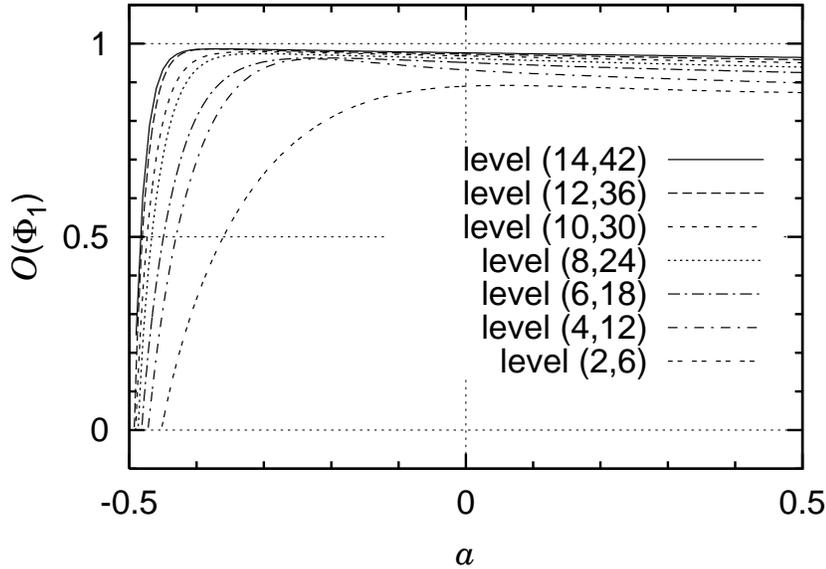}
\end{center}
\caption{Gauge invariant overlap for the numerical stable solution in the
 expanded theory.}
\label{fig:ginv}
\end{figure}
\begin{figure}[t]
 \begin{center}
\includegraphics[width=11.5cm]{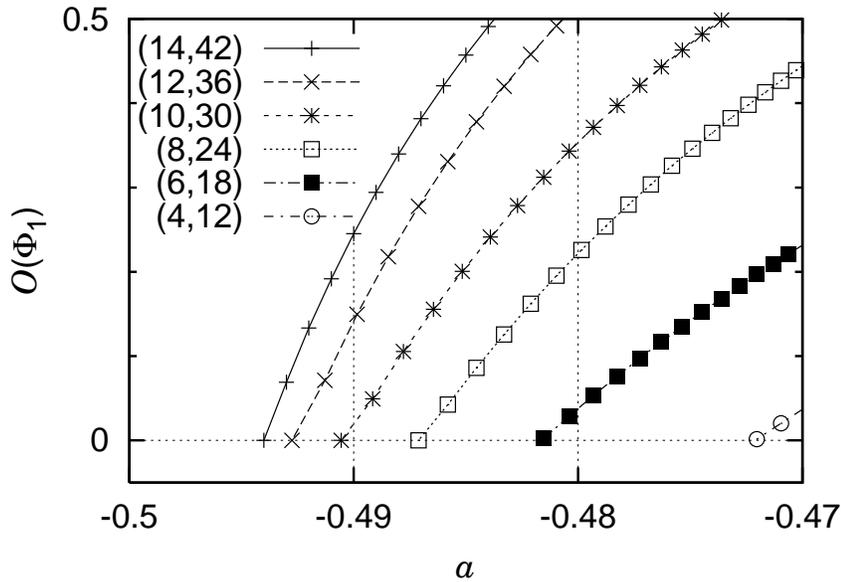}
\end{center}
\caption{Enlarged view of the gauge invariant overlap for $\Phi_1$
 around $a=-0.5$.}
\label{fig:ginv05}
\end{figure}

The plots of the
overlap are displayed in Fig.~\ref{fig:ginv} and the enlarged view
around $a=-1/2$ is shown in Fig.~\ref{fig:ginv05}. 
We find that the resulting plots approach the expected step function
as the truncation level is increased.
In the enlarged graph, 
we can find the point at which the solution becomes trivially zero
more clearly than in the vacuum energy case. This critical
point approaches $a=-1/2$ as the level is increased.

Hence, all the numerical results for the annihilation of the stable
solution confirm that the expanded theory has the vacuum structure as
depicted in Fig.~\ref{fig:vacenergy1}. 
In this  section, we checked the existence of the stable solution for
$a>-1/2$ and it annihilates at $a=-1/2$. 
\section{Emergence of unstable vacuum
\label{sec:emergence}}

In this section, we consider an unstable solution in the theory expanded
around the identity-based solution.
For $a>-1/2$, the expanded theory is 
unstable at $\Phi=0$. 
However, the expanded theory is expected to be
already on the stable vacuum for $a=-1/2$; namely, $\Phi=0$ is a stable
vacuum at $a=-1/2$. If that is the case, 
the expanded theory for $a=-1/2$ should have a nontrivial unstable
solution corresponding to the perturbative vacuum in the
original theory. 

\begin{figure}[b]
\begin{center}
\includegraphics[width=7cm]{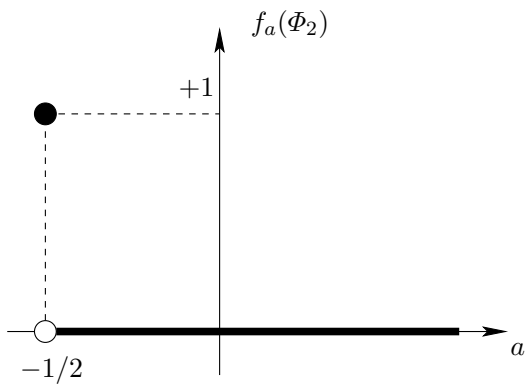}
\end{center}
\caption{Vacuum energy expected for the unstable solution
 $\Phi_2$.}
\label{fig:vacenergy4}
\end{figure}

Since the unstable solution $\Phi_2$ should correspond to the
perturbative open string vacuum, the vacuum energy of the unstable solution
should be equal to the D-brane tension (not the {\it minus} D-brane
tension). Therefore, the vacuum energy of $\Phi_2$
is expected to behave as depicted in
Fig.~\ref{fig:vacenergy4}. $f_a(\Phi_2)$ is trivially zero for $a>-1/2$,
but it should be increased to $+1$ 
owing to the emergence of the unstable solution at $a=-1/2$.
Similarly, we expect that the gauge invariant overlap for $\Phi_2$
should be $-1$ at $a=-1/2$ and trivially zero for $a>-1/2$.

Now, let us find the unstable solution $\Phi_2$ using level
truncation calculation. 
We also apply the iterative approximation algorithm
as used to find the stable solution in \S\ref{sec:annihilation}.
At $a=-1/2$, as the initial configuration $\Phi^{(0)}$, we take
\begin{eqnarray}
 \Phi^{(0)}=-\frac{32}{27\sqrt{3}}\,c_1\left|0\right>,
\end{eqnarray}
which is the nontrivial solution in the level $(0,0)$ truncation.
By the iteration (\ref{eq:iteration_La}), we obtain
a nontrivial solution $\Phi_2|_{a=-1/2}$.
Furthermore, for $a=-1/2+\epsilon$ $(0<\epsilon\ll 1)$, with this
solution $\Phi_2|_{a=-1/2}$ as the initial configuration, we can
construct a solution $\Phi_2|_{a=-1/2+\epsilon}$
using the iteration (\ref{eq:iteration_La}).
In the same manner, for $a=-1/2+(m+1)\epsilon$, we can uniquely 
construct a solution $\Phi_2|_{a=-1/2+(m+1)\epsilon}$
with the initial configuration  $\Phi_2|_{a=-1/2+m\epsilon}$
$(m=0,1,2,\cdots)$.
We find that $\Phi_2|_{a}$ reaches the trivial configuration
for some $m_0$: $\Phi_2|_{a=-1/2+m_0\epsilon}=0$.
Therefore, for $a\ge -1/2+m_0\epsilon$, the nontrivial unstable solution
$\Phi_2$ vanishes.

As in \S\ref{sec:annihilation},  we continue the iterative procedure
until the relative error reaches $10^{-8}$ and
examine whether the final configuration is truly physical in terms of
its BRST invariance.
We find that the solutions $\Phi_2$ constructed as above are 
consistent with the equation of motion $Q'\Phi+\Phi*\Phi=0$
with increasing level.

The vacuum energy and gauge invariant overlap for $\Phi_2$ at
$a=-1/2$ are shown in Table.~\ref{tab:vacginvtab}.\footnote{The vacuum
energy of the unstable solution at $a=-1/2$ was firstly written in
Ref.~\citen{rf:zeze} up to level 6. It was also discussed
in the context of vacuum string field theory also  up to level
6.\cite{Drukker:2005hr}} We find that the vacuum energy approaches
the expected value of $+1$ as 
the truncation level is increased. 
At level $(16,48)$, the vacuum energy is about 24\% over, 
although it is about 260\% at level $(2,6)$.
Moreover, the gauge invariant overlap also takes around the expected
value of $-1$. These results suggest that the level truncation
approximation is also applicable to the analysis of the unstable
solution.
Although it is not so efficient compared with that of the stable
solution, it is reasonably confirmed that the unstable solution
does exist as expected in the expanded theory at $a=-1/2$.
\renewcommand{\arraystretch}{.9}
\begin{table}[t]
\caption{Vacuum energy and gauge invariant overlap for the unstable
 solution $\Phi_2$ at $a=-1/2$.}
\label{tab:vacginvtab}
\begin{center}
 \begin{tabular}{|c|c|c|}
\hline
level&\makebox[4cm]{vacuum energy} & \makebox[4cm]{gauge inv. overlap}\\
\hline
\hline
$(0,0)$ & $2.3105795$  
  & $-1.0748441$\\ 
\hline
$(2,6)$ & $2.5641847$  
  & $-1.0156983$\\ 
\hline
$(4,12)$ & $1.6550774$  
  & $-0.9539832$\\ 
\hline
$(6,18)$ &  $1.6727496$ 
  & $-0.9207572$\\ 
\hline
$(8,24)$ & $1.4193393$ 
  & $-0.9377548$\\ 
\hline
$(10,30)$ & $1.4168893$ 
   & $-0.9110994$\\ 
\hline
$(12,36)$ & $1.3035715$ 
   & $-0.9237917$\\ 
\hline
$(14,42)$ & $1.2986472$ 
   & $-0.9056729$\\ 
\hline
$(16,48)$ & $1.2357748$ 
         & $-0.9229035$\\ 
\hline \end{tabular}\\
\end{center}
\end{table}

Note that the vacuum energy decreases to $+1$ with the
period of level 4.
This is a contrasting fact to the monotonically decreasing
tachyon vacuum energy in the original theory.
Probably, this behavior is considered to be related to the fact that
the kinetic operator (\ref{Eq:La}) mixes states with
level 2 difference.

In Ref.~\citen{Gaiotto:2002wy}, the vacuum energy up
to level $(18,54)$ was extrapolated to a higher level on the basis of
the study of an effective potential\cite{Taylor:2002fy} in the level
truncated theory. In fact, in the original theory ($a=0$), 
if we fit the stable vacuum energy $f_{a=0}(\Phi_1)$
by a function of the level $L$ such as
\begin{eqnarray}
\label{eq:fitfun}
 F_N(L)=\sum_{n=0}^N\frac{a_n}{(L+1)^n}
\end{eqnarray}
with the data for the level $(L,3L)$ $(L=0,2,4,6,8,10,12,14,16;N=9)$,
we find that the `straightforward' extrapolation gives the
vacuum energy at $L=\infty$ as
$\tilde E_{\infty}^{(16)}=F_9(\infty)=-1.00003$.\cite{Gaiotto:2002wy}
Similarly, at $a=-1/2$, let us consider a fit of the unstable vacuum
energy $f_{a=-1/2}(\Phi_2)$ using $F_N(L)$
(1) with the data for the level $(L,3L)$ $(L=0,4,8,12,16;N=5)$
and (2) with the data for the level $(L,3L)$ $(L=2,6,10,14;N=4)$.
Then, the straightforward extrapolation gives the
vacuum energy at $L=\infty$ as 
(1) $F_5(\infty)=0.98107$
and (2) $F_4(\infty)=0.98146$, respectively.
Both of them are $98\%$ of the expected value, which seems to be
an encouraging result although this extrapolation may lack a rigorous
justification.

We should comment on the increase in the gauge invariant overlap beyond
the expected value of $-1$.
Because we do not have a reasonable fitting method 
for the gauge invariant overlap, we have to examine higher level
behavior to clarify whether the gauge invariant overlap is away from
the expected value.

\begin{figure}[h]
 \begin{center}
\includegraphics[width=11.5cm]{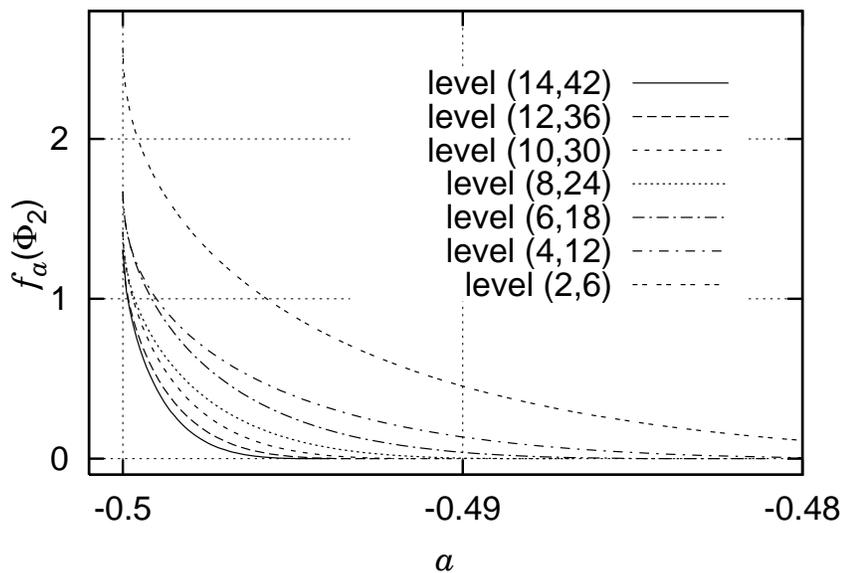}
\end{center}
\caption{Vacuum energy of the unstable solution for various $a$.}
\label{fig:unstablevac}
\end{figure}

\begin{figure}[h]
 \begin{center}
\includegraphics[width=11.5cm]{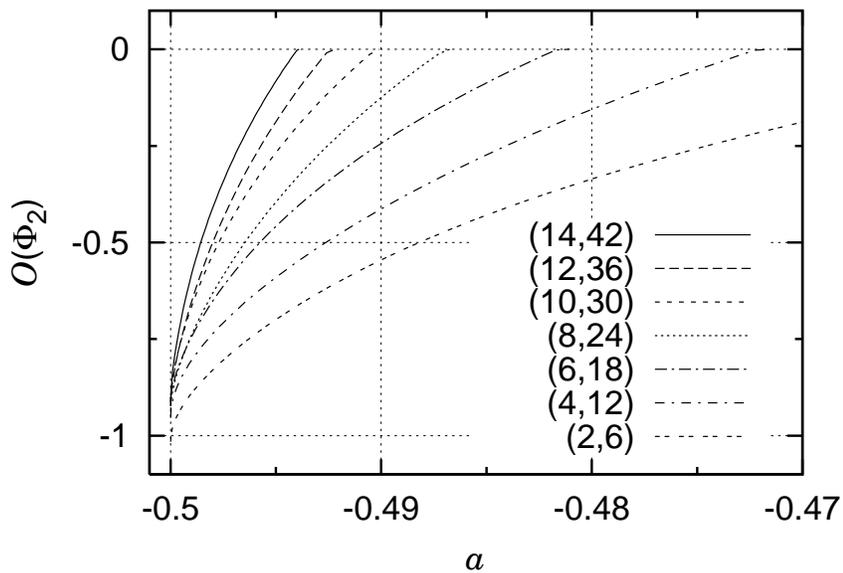}
\end{center}
\caption{Gauge invariant overlap of the unstable solution for various
 $a$.}
\label{fig:unstableginv}
\end{figure}

Finally, let us consider the unstable solution $\Phi_2$
for various $a\,(>-1/2)$.
The resulting vacuum energy of the unstable solution is depicted in
Fig.~\ref{fig:unstablevac}. The vacuum energy is around the
expected value for $a=-1/2$, but it decreases rapidly to zero for
increasing $a$.\footnote{Note that in
Fig.~\ref{fig:unstablevac} the parameter $a$ ranges from $-0.5$ to
$-0.48$ and this range is much narrower than that of
Fig.~\ref{fig:vacenergy2}.} This zero vacuum energy is due to the fact
that the solution itself becomes a trivial configuration $(\Phi_2=0)$ for
the parameter $a$ over a critical value. 
For example, at level 14, the
solution converges trivially into zero at about $a=-0.494$. This
critical value approaches 
$a=-1/2$ as the truncation level is increased. We can also find that the
critical value is nearly identical to that of the stable solution, at
which the stable solution becomes trivial for decreasing $a$ as in the
previous section.\footnote{We find that the critical values for $\Phi_1$
and $\Phi_2$ exactly coincide with the level 0 approximation.}
Thus, we find that the vacuum energy of the
unstable solution approaches the step function as expected in
Fig.~\ref{fig:vacenergy4} for increasing truncation levels.
For the gauge invariant overlap of the unstable solution, the resulting
plots are displayed in Fig.~\ref{fig:unstableginv}. Similarly to the
vacuum energy, the overlap becomes increasingly closer to the expected
behavior for higher truncation level. 

Although the above results are encouraging for our conjecture,
we eventually need more higher level investigation 
to provide more reliable and quantitative results.\cite{rf:KTprogress}

\section{Summary and discussion
\label{sec:summary}}

We have found
that the stable and unstable solutions in the Siegel gauge,
$\Phi_1$ and $\Phi_2$, numerically exist in the theory expanded around
the identity-based solution. For these solutions, we have evaluated the
vacuum energy 
and the gauge invariant overlap in terms of level truncation
approximation up to level $(14,42)$.
As the truncation level is increased, the
plots of the two gauge invariants for the parameter $a$ become
remarkably closer to the behavior expected from the vacuum structure in
Fig.~\ref{fig:vac}. These results strongly  support our expectation
that the identity-based solution corresponds to a trivial pure gauge
configuration for $a>-1/2$, but it can be regarded as the tachyon vacuum
solution for $a=-1/2$.  

In this paper, we have considered the identity-based solution constructed
only by the specific function (\ref{Eq:hz}). However, it is possible to
construct many identity-based solutions for various functions.
Concerning the stable solution, it was found that the vacuum energy for
various functions behaves similarly to that for the function in this
paper.\cite{Takahashi:2003ppa,Zeze:2004yh,Igarashi:2005wh,Igarashi:2005sd}
To clarify the nature of the identity-based solution, we should study
unstable solutions further for the various functions. 
Moreover, it would be better to study the extrapolation of our analysis
to higher levels. In any case, it is necessary to compute several
quantities by higher level truncation.

Our results suggest that it may be possible to evaluate directly the
vacuum energy and the gauge invariant overlap of the identity-based
solution. While the results are encouraging, the direct calculation
remains one of the most difficult issues in string field theory.
This difficulty seems to come from the lack of a reasonable
regularization scheme for the identity-based solution.
However, the success of the numerical analysis is a characteristic
feature of the identity-based solution. 
In addition to the Schnabl-type solutions,
the identity-based solution seems to provide
complementary approaches to a deeper understanding of the string field
theory. 

In particular, for the identity-based solution, there is an interesting
possibility of understanding closed strings on the tachyon vacuum.
The expanded theory around the identity-based solution
provides a worldsheet picture of perturbative
amplitudes.\cite{Drukker:2002ct,Drukker:2003hh,Zeze:2004yh,Igarashi:2005wh} 
For $a>-1/2$, the worldsheet has boundaries and this is consistent with
the expectation that the solution corresponds to a trivial pure gauge
configuration. However, the boundary existing for $a>-1/2$ shrinks into
a point as $a$ approaches $-1/2$, and the worldsheet is given as a
closed surface without boundaries.
This fact also confirms that the identity-based solution at $a=-1/2$ can
be regarded as the tachyon vacuum where there are no open strings.
Consequently, we expect that the worldsheet picture in the expanded
theory clarifies the existence of closed strings on the tachyon vacuum
and, moreover,
it may provide a quantitative computational method of closed string
amplitudes via open string fields.

\section*{Acknowledgements}

The work of I.~K. was supported in part by a Special Postdoctoral 
Researchers Program at RIKEN and a Grant-in-Aid for Young Scientists
(\#19740155) from MEXT of Japan.
The work of T.~T. was supported in part by a Grant-in-Aid for 
Young Scientists (\#18740152) from MEXT of Japan.
The level truncation calculations based on {\sl Mathematica} were 
carried out partly on the computer {\it sushiki} at Yukawa Institute 
for Theoretical Physics in Kyoto University.

%


\end{document}